\documentclass[11pt]{article}
\usepackage{indentfirst}

\newcommand{\be}{\begin{equation}}
\newcommand{\ee}{\end{equation}}
\newcommand{\bea}{\begin{eqnarray}}
\newcommand{\eea}{\end{eqnarray}}

\newcommand{\tr}{{\rm tr }}

\newtheorem{theorem}{Theorem}

\begin{document}

\title{On the Quantumness of a Hilbert Space}

\author{Christopher A. Fuchs\medskip
\\
\small Communications Network Research Institute,
Dublin Institute of Technology
\\
\small Rathmines Road, Dublin 6, Ireland \smallskip
\\
and \smallskip
\\
\small Bell Labs, Lucent Technologies
\\
\small 600-700 Mountain Avenue, Murray Hill, New Jersey 07974, USA}

\date{}

\maketitle

\abstract{We derive an exact expression for the quantumness of a
Hilbert space (defined in C.~A. Fuchs and M.~Sasaki, Quant.\ Info.\
Comp. {\bf 3}, 377 (2003)), and show that in composite Hilbert spaces
the signal states must contain at least some entangled states in
order to achieve such a sensitivity. Furthermore, we establish that
the accessible fidelity for symmetric informationally complete signal
ensembles is equal to the quantumness. Though spelling the most
trouble for an eavesdropper because of this, it turns out that the
accessible fidelity is nevertheless easy for her to achieve in this
case: Any measurement consisting of rank-one POVM elements is an
optimal measurement, and the simple procedure of reproducing the
projector associated with the measurement outcome is an optimal
output strategy.}

\section{Introduction}

Memorable experiences sometimes happen in elevators.  I have had two
in my life:  This paper has to do with both.

The setting of the second was QIC '96 in Fuji-Hakone, Japan.  It was
the first time I met Alexander Holevo, to whom this paper is
dedicated. As we entered an elevator, Richard Jozsa gave Prof.\
Holevo a brief description of the then recent quantum
channel-capacity superadditivity result of Ref.~\cite{Hausladen96}.
Holevo---apparently not quite absorbing what he had just heard---said
something like, ``The issue of quantum channel capacity is very
tricky.  For instance, collective measurements on individual signals
can increase capacity. There is no classical analogue to this.  You
can read about the phenomenon in my 1979 paper; I will give you the
reference.'' When Holevo left the elevator, Jozsa and I looked at
each other in awe! Had he really known this effect so long before
Ref.~\cite{Hausladen96} and even before Ref.~\cite{Peres91}?  Sure
enough~\cite{Holevo79}.  And like so many of Holevo's great early
contributions to quantum information theory, it went essentially
unnoticed for many years.  He has always been a man ahead of his
time.

In contrast, ten years before that second experience, I met an odd
fellow in an elevator at the physics department of the University of
Texas.  All alone, with the doors shut, he looked at me with a crazed
look in his eyes and asked, ``What is energy?'' Feeling
uncomfortable, I turned my own eyes to the ground and was happy that
the doors soon opened. I slipped away, but regaining composure just
before the doors shut again, I replied, ``I don't know, that which
gravitates?'' A cheap answer!  But at least I had something to say.
Looking back over the years I thank my lucky stars he didn't ask a
tough question. He could have asked, ``What is Hilbert space?''

Associated with each quantum system is a Hilbert space.  In the case
of finite dimensional ones, it is commonly said that the dimension
corresponds to the number of distinguishable states a system can
``have.''  But what are these distinguishable states?  Are they
potential properties a system can possess in and of itself, much like
a cat's possessing the binary value of whether it is alive or dead?
If the Bell-Kochen-Specker theorem~\cite{Appleby03} has taught us
anything, it has taught us that these distinguishable states should
not be thought of in that way.

In this paper, I present some results that take their {\it
motivation\/} (though not necessarily their interpretation) in a
different point of view about the meaning of a system's
dimensionality. From this view, dimensionality may be the raw,
irreducible concept---the single {\it property\/} of a quantum
system---from which other consequences are derived (for instance, the
maximum number of distinguishable preparations which can be imparted
to a system in a communication setting)~\cite{Fuchs02,FuchsWHAT}. The
best I can put my finger on it is that dimensionality should have
something to do with a quantum system's ``sensitivity to the
touch,''~\cite{FuchsPaulian,Fuchs01} its ability to be modified with
respect to the external world due to the interventions of that world
upon its natural course.  Thus, for instance, in quantum computing
each little push or computational step has the chance of counting for
more than in the classical world.

Various aspects of quantum eavesdropping seem to be perfect for
sussing out and quantifying such ideas.  One is the setting
introduced in Ref.~\cite{Fuchs03} and explored further in
Ref.~\cite{Audenaert04}. Here I show that, of the definitions spelled
out there, the quantumness of a Hilbert space ${\cal H}_d$ of
dimension $d$ can be calculated explicitly. Moreover, if a certain
class of symmetric signal ensembles exists, such a sensitivity to
eavesdropping can by achieved by using signals drawn from an ensemble
of no more than $d^2$ elements.  Interestingly, for composite
systems, entangled states are a necessary ingredient for achieving
the quantumness:  In particular, by the measure of quantumness,
systems' ``sensitivity to the touch'' is strictly
supermultiplicative.

The plan of the paper is as follows.  In
Section~\ref{NotationsSection}, I reacquaint the reader with the
definitions of Ref.~\cite{Fuchs03}. In
Section~\ref{QuantumnessSection}, I derive an expression for the
quantumness of ${\cal H}_d$.  I also point out that on composite
Hilbert spaces ${\cal H}_{d_1}\otimes{\cal H}_{d_2}$, ensembles
containing entangled states are necessary for achieving the
quantumness. In Section~\ref{SICSection}, I introduce the idea of a
symmetric informationally complete signal ensemble~\cite{Renes03} and
show that---if it exists---it achieves the quantumness of the Hilbert
space.\footnote{In another connection, the use of such ensembles for
quantum cryptography has also been considered in
Refs.~\cite{Renes04a,Renes04b}.} In
Section~\ref{OtherEnsemblesSection}, I show that when a complete set
of mutually unbiased bases~\cite{Wootters89,Bandyopadhyay02} exists
for ${\cal H}_d$, signals drawn from such an ensemble of $d(d+1)$
elements also achieve the quantumness. Furthermore, I look into the
question of the minimal number of elements in an ensemble required
for it to achieve the quantumness of ${\cal H}_d$.  In
Section~\ref{OpenQuestionsSection}, I reemphasize the open question,
and finally in Section~\ref{ConcludingRemarks}, I conclude with the
hint of another elevator story.

\section{Preliminary Notations}
\label{NotationsSection}

Recall the main definitions from Ref.~\cite{Fuchs03}.  Given a signal
ensemble ${\cal P}$ (i.e., a collection ${\cal P}=\{\,\Pi_i,\pi_i\}$
of pure quantum states $\Pi_i=|\psi_i\rangle\langle\psi_i|$ along
with associated probabilities $\pi_i$), a measurement ${\cal
E}=\{E_b\}$ (i.e., a positive operator-valued measure or
POVM~\cite{Davies70}) and a state-reproduction strategy ${\cal
M}\!:b\rightarrow \sigma_b$ (i.e., a map taking measurement outcomes
to new quantum states), we can define an {\it average fidelity\/} for
${\cal E}$ and ${\cal M}$ according to
\be
F_{\cal P}({\cal E},{\cal M}) = \sum_{b,i} \pi_i \tr(\Pi_i E_b)
\tr(\Pi_i\sigma_b)\;,
\ee
The average fidelity represents an eavesdropper's probability of
going unnoticed after performing an ``intercept-resend'' strategy of
this type.

The {\it achievable fidelity\/} for a given measurement $\cal E$ is
the average fidelity optimized over all reconstruction strategies
$\cal M$:
\be
F_{\cal P}({\cal E})=\sup_{\cal M}F_{\cal P}({\cal
E},{\cal M})\;.
\ee
The achievable fidelity, it turns out, can be explicitly
calculated~\cite{Fuchs03} in terms of the trace-nonincreasing
completely positive linear map---the {\it ensemble map}---$\Psi:{\cal
L}({\cal H}_d)\rightarrow {\cal L}({\cal H}_d)$ defined by
\be
\Psi(X)=\sum_i \pi_i \Pi_i X \Pi_i\;.
\label{NextDayBirdsChirping}
\ee
With it,
\be
F_{\cal P}({\cal E})=\sum_b \lambda_1 \big(\Psi(E_b)\big)\;,
\label{FumbleThumbs}
\ee
where $\lambda_1(X)$ denotes the largest eigenvalue of a Hermitian
operator $X$.

The {\it accessible fidelity\/} of the ensemble $\cal P$ is the best
possible average fidelity an eavesdropper can attain over all
measurements and all reproduction strategies:
\bea
F_{\cal P} &=& \sup_{{\cal E},{\cal M}}F_{\cal P}({\cal E},{\cal M})
\nonumber\\
&=&
\sup_{{\cal E}}\sum_b \lambda_1 \big(\Psi(E_b)\big)\;.
\eea
Thus, the accessible fidelity is a natural measure of an ensemble's
intrinsic sensitivity to eavesdropping.

Finally, the {\it quantumness\/} of a Hilbert space ${\cal H}_d$ is
smallest possible accessible fidelity the space can support
\be
Q_d=\inf_{\cal P}F_{\cal P}\;.
\ee
The quantumness, it should be noted, is an inverted measure:  The
smaller the quantumness of a Hilbert space, the greater a system's
ultimate sensitivity to intercept-resend style quantum eavesdropping.

\section{Quantumness of a Hilbert Space}
\label{QuantumnessSection}

Of the various ensembles explored in Ref.~\cite{Fuchs03}, the one
with the smallest accessible fidelity was the (unique) unitarily
invariant ensemble on ${\cal H}_d$.  Its accessible fidelity was
shown to be $2/(d+1)$.  This establishes that
\be
Q_d\le\frac{2}{d+1}\;.
\ee
Since it is hard to imagine a more difficult ensemble for an
eavesdropper to successfully eavesdrop upon than this one, it was
speculated in Ref.~\cite{Fuchs03} that
\be
Q_d=\frac{2}{d+1}\;.
\label{LameBrain}
\ee
To prove this is the case, we establish the following theorem.
\begin{theorem}
For any ensemble $\cal P$, the accessible fidelity for that ensemble
satisfies
\be
F_{\cal P}\ge\frac{2}{d+1}\;.
\ee
\end{theorem}

To see this, we use a trick similar to the one used in
Ref.~\cite{Jozsa94}. For any (discrete) ensemble ${\cal
P}=\{\Pi_i,\pi_i\}$, imagine an eavesdropper partaking in the
following strategy.  She performs a standard, but random, von Neumann
measurement ${\cal G}=\{G_b\}_{b=1}^d$ consisting of one-dimensional
projection operators, and then uses a strategy $\cal M$ that simply
reproduces the state $G_b$ corresponding to the outcome she finds. By
definition,
\be
F_{\cal P}\ge F_{\cal P}({\cal G},{\cal M})=\sum_{b,i}\pi_i \big(\tr
\Pi_i G_b\big)^2\;.
\ee
However, also,
\be
F_{\cal P}\ge \overline{\,F_{\cal P}({\cal G},{\cal M})\,}\;,
\ee
where the overline represents an average over all such measurements,
for instance with respect to the unitarily invariant measure.

Thus, all we need to do is resurrect the measure $d\Omega_\psi$ of
Eq.\ (111) in Ref.~\cite{Fuchs03}, along with the result of Eq.\
(114) there, to get:
\bea
\overline{\,F_{\cal P}({\cal G},{\cal M})\,}
&=&
\sum_i \sum_{b=1}^d \pi_i \int \big(\tr \Pi_i G_b\big)^2
d\Omega_{\cal G} \; =\; d \sum_i \pi_i \int \big(\tr \Pi_i G\big)^2
d\Omega_G
\nonumber\\
&=&
d \sum_i \pi_i \int |\langle\psi_i|\phi\rangle|^4 d\Omega_\phi \; =\;
d \sum_i \pi_i \frac{\Gamma(d)\Gamma(3)}{\Gamma(1)\Gamma(d+2)}
\nonumber\\
&=&
d \frac{(d-1)!\,2!}{(d+1)!} \;=\; \frac{2}{d+1}
\eea
Here, $d\Omega_{\cal G}$, $d\Omega_G$, and $d\Omega_\phi$ represent
the various incarnations of the unitarily invariant measure as it is
translated from being about complete von Neumann measurements to
projection operators on ${\cal H}_d$ to normalized vectors in ${\cal
H}_d$, respectively, and $G=|\phi\rangle\langle\phi|$ is a dummy
one-dimensional projector.

It is interesting to couple Eq.~(\ref{LameBrain}) with the findings
of Ref.~\cite{Audenaert04}.  By the result here, if we consider a
composite Hilbert space ${\cal H}_{d_1}\otimes{\cal H}_{d_2}$, with
components of dimension $d_1$ and $d_2$, its quantumness is given by
\be
Q_{\rm comp}=\frac{2}{\,d_1 d_2 + 1\,}\;.
\ee
On the other hand, in Ref.~\cite{Audenaert04} a general
multiplicativity result was proven for sets of product states.  That
result along with Eq.~(\ref{LameBrain}) shows that the smallest
accessible fidelity that can be achieved with signal ensembles of
product states is
\be
F_{{\cal P}_1\otimes{\cal
P}_2}=\left(\frac{2}{d_1+1}\right)\!\!\left(\frac{2}{d_2+1}\right)\;.
\ee
The implication of this is that ensembles $\tilde{\cal P}$ optimal
for achieving the quantumness of ${\cal H}_{d_1}\otimes{\cal
H}_{d_2}$ must contain entangled states.  Indeed the quantumness of a
composite Hilbert space is not simply multiplicative in the
quantumnesses of its components.

\section{Symmetric Informationally Complete Ensembles}
\label{SICSection}

{\it Suppose\/}\footnote{Be careful to note that this is no trivial
supposition.  The problem of the existence of such a set, in fact,
has existed in the mathematical literature since the early
1970s~\cite{Lemmens73,Delsarte75}.  To date, such sets have only been
proven to exist in dimensions
$d=2,3,4,8$~\cite{Renes03,Hoggar98,CavesNotes}. The remainder of the
evidence for their existence (in dimensions up to $d=45$) comes
through numerical work~\cite{Renes03}.  Thus, it does seem likely
that such sets exist, but it cannot be taken for granted.} there
exists $d^2$ unit vectors $|\psi_i\rangle\in{\cal H}_d$ such that
\be
|\langle\psi_i|\psi_j\rangle|^2=\frac{1}{d+1}\quad\forall i\ne j\;.
\label{RushLimbaugh}
\ee
If such a set exists, it follows that the $d^2$ projection operators
$\Pi_i= |\psi_i\rangle\langle\psi_i|$ form a linearly independent
set~\cite{CavesNotes}.  To see this, suppose there exist real numbers
$\alpha_i$ such that
\be
\sum_i \alpha_i \Pi_i = 0\;.
\label{first}
\ee
Multiplying by $\Pi_k$ and taking the trace of both sides we get
\be
\alpha_k + \frac{1}{d+1}\sum_{i\ne
k}\alpha_i=\left(1-\frac{1}{d+1}\right)\alpha_k+\frac{1}{d+1}\sum_i\alpha_i=0\;,
\ee
which implies that, for all $\alpha_k$,
\be
\alpha_k=-\frac{1}{d}\sum_i\alpha_i\;.
\ee
On the other hand, taking the trace of Eq.~(\ref{first}) reveals that
$\sum_i\alpha_i=0$.  It follows that all $\alpha_k=0$ for all $k$.
The $\Pi_i$ are thus linearly independent.

Because of this latter property, the projectors $\Pi_i$ form a
complete basis on ${\cal L}({\cal H}_d)$, the vector space of linear
operators over ${\cal H}_d$.  It follows that for any operator
$X\in{\cal L}({\cal H}_d)$, the $d^2$ numbers $\tr X \Pi_i$ generated
by the Hilbert-Schmidt inner product $(A,B)=\tr A^\dagger B$ uniquely
specify the operator $X$.

It is convenient to use the projectors $\Pi_i$ to form a completely
positive linear map $\Phi:{\cal L}({\cal H}_d)\rightarrow {\cal
L}({\cal H}_d)$ in the following way:
\be
\Phi(X)=\sum_i\frac{1}{d^2}\Pi_iX\Pi_i\;.
\label{second}
\ee

\begin{theorem}
An alternative representation of $\Phi$ is this:
\be
\Phi(X)=\frac{1}{d(d+1)}\Big((\tr X)I+ X\Big)\;,
\label{HumVee}
\ee
where $I$ denotes the identity operator.
\end{theorem}

To see this, note that $\Phi(I)$ is a density operator and that
\be
\tr\Big(\Phi(I)^2\Big)=\frac{1}{d}\;.
\ee
Thus,
\be
\Phi(I)=\frac{1}{d}I\;.
\ee
Now, for any $X\in {\cal L}({\cal H}_d)$ there exists an expansion
\be
X=\sum c_i \Pi_i\;,
\ee
where
\be
\tr X = \sum_i c_i\;.
\ee
With these ingredients, simply follow the action of $\Phi$ on $X$:
\bea
\Phi(X) &=& \frac{1}{d^2}\sum_{ij} c_j \Pi_i\Pi_j\Pi_i
\nonumber
\\
&=& \frac{1}{d^2}\sum_{ij} c_j \tr(\Pi_i\Pi_j)\Pi_i
\nonumber
\\
&=& \frac{1}{d^2}\sum_{i} c_i \Pi_i +
\frac{1}{d^2(d+1)}\sum_{i\ne j} c_j \Pi_i
\nonumber
\\
&=& \frac{1}{d^2}X + \frac{1}{d^2(d+1)}\left(
\sum_{ij} c_j \Pi_i-\sum_{i} c_i \Pi_i\right)
\nonumber
\\
&=& \frac{1}{d^2}X + \frac{1}{d^2(d+1)}\left(
\Big(\sum_{j} c_j\Big)\Big(\sum_i \Pi_i\Big)-X\right)
\nonumber
\\
&=& \frac{1}{d^2}X + \frac{1}{d^2(d+1)}\Big(
d(\tr X)I-X\Big)
\nonumber
\\
&=& \frac{1}{d(d+1)}\Big((\tr X)I+X\Big)\;.
\label{fourth}
\eea
This proves the theorem.

When $\Phi$ acts on a density operator $\rho$, its action is (up to a
scaling factor) that of a ``just barely'' entanglement-breaking
depolarizing channel. In the language of Ref.~\cite{King02},
\be
\Phi(\rho)=\frac{1}{d}\Delta_\lambda(\rho)\;,
\qquad\mbox{with}\qquad\lambda=\frac{1}{d+1}\;.
\ee
Also note that by acting $\Phi$ on the identity, we obtained that
$\sum \frac{1}{d} \Pi_i = I$, which means the operators
\be
E_i=\frac{1}{d}\Pi_i
\ee
form a positive operator-valued measure.  Such a POVM is known as a
symmetric informationally complete POVM, or SIC-POVM for
short~\cite{Renes03}. The appellation `informationally complete' is
used because for any density operator $\rho$, if one knows the
probabilities
\be
p(i)=\tr\rho E_i
\ee
for the outcomes of such a measurement, then one knows the operator
$\rho$ itself~\cite{Caves02,DAriano03}.  In fact, using
Eq.~(\ref{HumVee}), one sees immediately that for any density
operator $\rho$
\be
\rho=(d+1)\sum_i p(i)\Pi_i - I\;.
\label{TheFormula}
\ee
For all these reasons, we will call any ensemble
\be
{\cal P}=\left\{\Pi_i,\frac{1}{d^2}\right\}
\ee
satisfying Eq.~(\ref{RushLimbaugh}), a {\it symmetric informationally
complete ensemble\/} (or SIC ensemble for short) in analogy to the
SIC-POVMs studied in Ref.~\cite{Renes03}.

Another useful quantity to know in these terms is the purity of
$\rho$ :
\bea
\tr\rho^2 &=& d^2(d+1)^2\sum_{gh}p(g)p(h)\tr E_gE_h - 2d(d+1)\sum_h
p(h)\tr E_h + \tr I
\nonumber\\
&=&
d^2(d+1)^2\sum_{gh}p(g)p(h)\tr E_gE_h - 2\big(\tr\rho+\tr I\big) +
\tr I
\nonumber\\
&=&
(d+1)^2\sum_h p(h)^2 + (d+1)\sum_{g\ne h}p(g)p(h) - d - 2
\nonumber\\
&=&
\left((d+1)^2-(d+1)\right)\sum_h p(h)^2 + (d+1)\sum_{g, h}p(g)p(h) -
d - 2
\nonumber\\
&=&
d(d+1)\sum_h p(h)^2-1
\eea
Thus all pure states give rise to a probability distribution $p(h)$
for the outcomes of a SIC-POVM such that
\be
\sum_h p(h)^2 = \frac{2}{d(d+1)}\;.
\label{BirdsChirping}
\ee

With these preliminary remarks, we are ready to explore the
accessible fidelity for SIC-ensembles.

\begin{theorem}
The accessible fidelity for any SIC ensemble $\cal P$ is given by
\be
F_{\cal P}=\frac{2}{d+1}\;,
\ee
and so achieves the quantumness of the Hilbert space. Moreover, any
POVM $\{G_b\}$ consisting of rank-1 elements $G_b=g_b
|\phi_b\rangle\langle\phi_b|$ can be used for an optimal
eavesdropping strategy.
\end{theorem}

To prove this, we simply fix any POVM ${\cal G}=\{G_b\}$ consisting
of rank-1 elements and use the general formula derived in for the
achievable fidelity in Eq.~(\ref{FumbleThumbs}):
\be
F_{\cal P}({\cal G})=\sum_b\lambda_1\big(\Phi(G_b)\big)\;.
\ee
With this,
\bea
F_{\cal P}({\cal G})
&=&
\sum_b g_b\lambda_1\!\!\left(\frac{1}{d(d+1)}\big(I+
|\phi_b\rangle\langle\phi_b|\big)\right)
\nonumber
\\
&=&
\frac{2}{d(d+1)}\sum_b g_b\;.
\eea
Finally, because $I=\sum_b G_b$, it follows that $\sum_b g_b=d$.
Thus,
\be
F_{\cal P}({\cal G})=\frac{2}{d+1}
\ee
regardless of the measurement $\cal G$ (so long as it consists of
rank-1 elements).  In particular,
\be
F_{\cal P}=\frac{2}{d+1}\;.
\ee

Furthermore notice that $F_{\cal P}$ can be achieved through a very
simple reconstruction strategy $\cal M$.  In particular, we do not
need to use the more difficult-to-express measurement derived in
Ref.~\cite{Fuchs03} which gives rise to Eq.~(\ref{FumbleThumbs}).

\begin{theorem}
For any measurement consisting of rank-1 elements $G_b= g_b
|\phi_b\rangle\langle\phi_b|\equiv g_b \sigma_b$, the accessible
fidelity $F_{\cal P}$ can be achieved via the simple reconstruction
strategy ${\cal M}\!:b\rightarrow\sigma_b$.\footnote{A theorem like
this was first shown for the case of the unitarily invariant signal
ensemble by Barnum in Ref.~\cite{Barnum02}.}
\end{theorem}

To prove this, define the conditional probabilities $p(i|b)$ by
\be
p(i|b)=\frac{1}{d}\,\tr\big(\Pi_i|\phi_b\rangle\langle\phi_b|\big)=
\tr\big(|\phi_b\rangle\langle\phi_b|E_i\big)\;,
\label{MidAtlantic}
\ee
Now, simply write out the expression for the average fidelity of such
a strategy.
\bea
F_{\cal P}({\cal G},{\cal M}) &=& \sum_{b,i} \frac{1}{d^2} \tr(\Pi_i
G_b) \tr(\Pi_i\sigma_b)
\nonumber
\\
&=& \sum_{b,i}  g_b \big(\tr(E_i
\sigma_b)\big)^2
\eea
Using the conditional probabilities in Eq.~(\ref{MidAtlantic}), this
becomes
\be
F_{\cal P}({\cal G},{\cal M})=\sum_b g_b \sum_i p(i|b)^2\;.
\ee
Noting that the $\sigma_b$ are pure states, so that
Eq.~(\ref{BirdsChirping}) is satisfied for the conditional
probabilities, we have finally
\bea
F_{\cal P}({\cal G},{\cal M}) &=& \frac{2}{d(d+1)}\sum_b g_b
\nonumber
\\
&=& \frac{2}{d+1}\;,
\eea
which is just the accessible fidelity.

\section{Other Ensembles Achieving Quantumness}
\label{OtherEnsemblesSection}

It is clear that for any other ensemble ${\cal P}=\{\,\Pi_i,\pi_i\}$,
if its ensemble map $\Psi$, Eq.~(\ref{NextDayBirdsChirping}), happens
to coincide with $\Phi$ in Eq.~(\ref{HumVee}), then that ensemble too
will have an accessible fidelity that achieves the quantumness of the
Hilbert space.  Here is another example.

Suppose ${\cal H}_d$ can be equipped with a complete set of mutually
unbiased bases.  That is, suppose one can find $d(d+1)$ one
dimensional projectors $\Pi^j_i$, with $j=1,\ldots,d+1$ and
$i=1,\ldots,d$, such that
\bea
\tr\big(\Pi^j_i\Pi^j_k\big) &=& \delta_{ik}
\\
\tr\big(\Pi^j_i\Pi^l_k\big) &=& \frac{1}{d} \quad\mbox{when}\quad
j\ne l\;.
\label{fifth}
\eea
It is known that such sets always exist when $d$ is an integer power
of a prime number~\cite{Wootters89,Bandyopadhyay02}.  (Though it is
speculated that they do not exist for general $d$, for instance, for
$d=6$~\cite{WoottersPrivate}.)  When such a set exists, it provides
an (overcomplete) basis for ${\cal L}({\cal H}_d)$.  Thus one can
write any operator $X$ in the form
\be
X=\sum_{ij} \alpha^j_i\, \Pi^j_i\;,
\ee
where the $\alpha^j_i$ are $d(d+1)$ complex numbers.  Performing now
a calculation similar to the one in Eq.~(\ref{fourth}), one obtains
that
\be
\sum_{ij} \frac{1}{d(d+1)}\,\Pi^j_i X \Pi^j_i = \Phi(X)\;.
\ee
Hence, an ensemble consisting of elements drawn from a complete set
of mutually unbiased bases (all equally weighted) achieves the
quantumness of the Hilbert space.

One can also ask the question of whether there are any ensembles with
strictly less than $d^2$ elements that still achieve the quantumness
of the Hilbert space.  If there are such ensembles, then it will have
to be for a reason more subtle than that the ensemble map $\Psi$ in
Eq.~(\ref{NextDayBirdsChirping}) coincides with $\Phi$. For, proving
$\Psi=\Phi$ is a sufficient condition achieving the quantumness, but
it is not a priori a necessary condition.

However, one can show the following about this sufficient condition:

\begin{theorem}
For each one-dimensional projector $\Pi\in{\cal L}({\cal H}_d)$,
define the completely positive linear map $\Phi_\Pi:{\cal L}({\cal
H}_d)\rightarrow {\cal L}({\cal H}_d)$ by
\be
\Phi_\Pi(X)=\Pi X\Pi\;.
\ee
Denote by $\cal Q$ the set of all such maps, and let $\cal B$ be the
convex hull of $\cal Q$.

If the map $\Phi$ in Eq.~(\ref{HumVee}) can be written as a convex
combination of $d^2$ or less extremal maps $\Phi_\Pi$ of $\cal B$,
then the projectors $\Pi$ in such a decomposition of $\Phi$ must
correspond to a SIC ensemble.
\end{theorem}

Here is how to see this.  Let $\{\Pi_i\}$ be the set projectors in
such a decomposition.  Note that there must be $d^2$ of them and that
they must be linearly independent.  This follows for the simple
reason that the range of $\Phi$ spans ${\cal L}({\cal H}_d)$. If some
of the $\Pi_i$ were linearly dependent or there were less than $d^2$
of them, then, for any probability distribution $\pi_i$, the
operators
\be
\sum_i \pi_i \Pi_i X \Pi_i= \sum_i \alpha_i \Pi_i\quad \forall X\;,
\ee
where $\alpha_i = \pi_i \langle\psi_i|X|\psi_i\rangle$, will not be
able to span ${\cal L}({\cal H}_d)$.

Now, working with the fact that the $\Pi_i$ are linearly independent,
try to satisfy the two equations:
\be
\sum_i \pi_i \Pi_i I \Pi_i = \Phi(I)= \frac{1}{d} I
\ee
and
\be
\sum_i \pi_i \Pi_i \Pi_k \Pi_i = \Phi(\Pi_k)=
\frac{1}{d(d+1)}\Big(I+\Pi_k\Big)\;.
\ee
Putting these two equation together, one obtains
\be
\left(d \pi_k -\frac{1}{d}\right)\Pi_k + (d+1)\sum_{i\ne k} \pi_i
\!\left( \tr(\Pi_i\Pi_k)-\frac{1}{d+1}\right)\Pi_i=0.
\ee
By the linear independence of the $\Pi_i$, the only way to satisfy
this is to have
\be
\pi_i = \frac{1}{d^2} \quad \forall i\;,
\ee
and
\be
\tr(\Pi_i\Pi_j)=\frac{1}{d+1}\;,\quad\forall i\ne j\;.
\ee
That completes the proof.

\section{Open Question}
\label{OpenQuestionsSection}

The previous section still leaves the open question:  Are there any
ensembles ${\cal P}=\{\,\Pi_i,\pi_i\}$ with strictly less than $d^2$
elements such that
\be
F_{\cal P}=Q_d\;?
\ee
If there are, what are the minimal number of elements required of an
ensemble so that it achieves the quantumness of the Hilbert space?

Whatever the answer---whether it be $d^2$ or strictly less---what is
the essential structure of such sets?  The suspicion here is that
this structure will have more to do with the intrinsic defining
characteristics of a quantum system than anything based on the
imagery of ``the number of distinguishable states a system can
have.''

\section{Concluding Remarks}
\label{ConcludingRemarks}

In several pieces of recent literature much to-do has been made of
the fungibility of quantum information~\cite{Caves03}.  Hilbert
spaces of the same dimension are said to be fungible: What can be
done in one can be done in the others.  Thus, for instance, it would
not matter if a quantum cryptographer decided to build a quantum key
distribution scheme based on $d$-dimensional subspaces gotten from
pellets of platinum, or $d$-dimensional subspaces gotten from pellets
of magnalium.  The ultimate security he can achieve---at least in
principle---will be the same in either case.

What is the meaning of this?  One might say that this is the very
reason quantum information is quantum {\it information\/}!  If a
protocol like quantum key distribution depended upon the kinds of
matter used for its implementation, one would hardly be justified in
thinking of it as a pure protocol solely of (quantum)
information-theoretic origin.  That is a useful and fruitful point of
view.

However, another point of view is that this may be a call to
reexamine physics, much like the miraculous equivalence between
gravitational and inertial mass (as revealed by the E\"otv\"os
experiment~\cite{Eotvos22}) was once a call to reexamine the origin
of gravitation.  From the perspective of gravity, the
``implementation'' of the mass is inconsequential---platinum and
magnalium fall with the same acceleration. And that, in the hands of
Einstein, led ultimately to the realization that gravity is a
manifestation of spacetime curvature.

What is Hilbert space?  Who knows!  But in quantum information we are
learning how to make use of it as a raw resource, basking in the good
fortune that it {\it is\/} fungible---that the implementation of a
Hilbert space dimension $d$ is inconsequential. There could be some
very deep physics in that, but it might take another elevator story.

\section*{Acknowledgements}

This paper is dedicated to Alexander Holevo on the occasion of his
60th birthday.  I thank Osamu Hirota, Chris King, Masahide Sasaki,
and Peter Shor for useful discussions, and especially Greg Comer for
keeping me close to gravity. I thank Ken Duffy for teaching me about
magnalium.


\begin{thebibliography}{99}

\bibitem{Hausladen96}
P.~Hausladen, R.~Jozsa, B.~Schumacher, M.~Westmoreland, and W.~K.
Wootters, ``Classical Information Capacity of a Quantum Channel,''
Phys.\ Rev.\ A {\bf 54}, 1869--1876 (1996).

\bibitem{Peres91}
A.~Peres and W.~K. Wootters, ``Optimal Detection of Quantum
Information,'' Phys.\ Rev.\ Lett.\ {\bf 66}, 1119--1122 (1991).

\bibitem{Holevo79}
A.~S. Holevo, ``Capacity of a Quantum Communication Channel,'' Prob.\
Info.\ Trans.\ {\bf 15}, 247--253 (1979).

\bibitem{Appleby03}
See for instance D.~M. Appleby, ``The Bell-Kochen-Specker Theorem,''
{\tt quant-ph/0308114}, and references therein.

\bibitem{Fuchs02}
C.~A. Fuchs, ``Quantum Mechanics as Quantum Information (and only a
little more),''  {\tt quant-ph/0205039}.

\bibitem{FuchsWHAT}
C.~A. Fuchs, ``Quantum States: What the Hell Are They?,'' posted at
{\tt http://netlib. bell-labs.com/who/cafuchs}.

\bibitem{FuchsPaulian}
C.~A. Fuchs, {\sl Notes on a Paulian Idea:  Foundational, Historical,
Anecdotal \& Forward-Looking Thoughts on the Quantum}, with foreword
by N. David Mermin, (V\"axj\"o University Press, V\"axj\"o, Sweden,
2003); see also {\tt quant-ph/0105039}.

\bibitem{Fuchs01}
C.~A. Fuchs and K.~Jacobs, ``Information-Tradeoff Relations for
Finite-Strength Quantum Measurements,'' Phys.\ Rev.\ A  {\bf 63},
062305 (2001).

\bibitem{Fuchs03}
C.~A. Fuchs and M.~Sasaki, ``Squeezing Quantum Information through a
Classical Channel:\ Measuring the `Quantumness' of a Set of Quantum
States,'' Quant.\ Info.\ Comp. {\bf 3}, 377--404 (2003); see also
{\tt quant- ph/0302092}.

\bibitem{Audenaert04}
K.~M.~R. Audenaert, C.~A. Fuchs, C.~King, and A.~Winter,
``Multiplicativity of Accessible Fidelity and Quantumness for Sets of
Quantum States,'' Quant.\ Info.\ Comp. {\bf 4}, 1--11 (2004); see
also {\tt quant-ph/0308120}.

\bibitem{Renes03}
J.~M. Renes, R.~Blume-Kohout, A.~J. Scott, C.~M. Caves, ``Symmetric
Informationally Complete Quantum Measurements,'' to appear in J.\
Math.\ Phys.\ (2004); see also {\tt quant-ph/0310075}.

\bibitem{Renes04a}
J.~M. Renes, ``Quantum Key Distribution Using Equiangular Spherical
Codes,'' {\tt quant-ph/0311106}.

\bibitem{Renes04b}
J.~M. Renes, ``Spherical Code Key Distribution Protocols for
Qubits,'' {\tt quant-ph/0402135}.

\bibitem{Wootters89}
W.~K. Wootters and B.~D. Fields, ``Optimal State-Determination by
Mutually Unbiased Measurements,'' Ann.\ Phys.\ {\bf 191}, 363--381
(1989).

\bibitem{Bandyopadhyay02}
S.~Bandyopadhyay, P.~O. Boykin, V.~Roychowdhury, and F.~Vatan, ``A
New Proof for the Existence of Mutually Unbiased Bases,''
Algorithmica {\bf 34}, 512--528 (2002); see also {\tt
quant-ph/0103162}.

\bibitem{Davies70}
E.~B. Davies and J.~T. Lewis, ``An Operational Approach to Quantum
Probability,'' Comm.\ Math.\ Phys.\ {\bf 17}, 239--260 (1970).

\bibitem{Jozsa94}
R.~Jozsa, D.~Robb, and W.~K. Wootters, ``Lower Bound for Accessible
Information in Quantum Mechanics,'' Phys.\ Rev.\ A {\bf 49}, 668--677
(1994).

\bibitem{Lemmens73}
P.~W.~H. Lemmens and J.~J. Seidel, ``Equiangular Lines,'' J.\ Algebra
{\bf 24}, 494--512 (1973).

\bibitem{Delsarte75}
P.~Delsarte, J.~M. Goethels, and J.~J. Seidel, ``Bounds for Systems
of Lines and Jacobi Polynomials,'' Philips Research Reports {\bf 30},
91--105 (1975).

\bibitem{Hoggar98}
S.~G. Hoggar, ``64 Lines from a Quaternionic Polytope,'' Geometriae
Dedicata {\bf 69}, 287--289 (1998).

\bibitem{CavesNotes}
C.~M. Caves, ``Symmetric Informationally Complete POVMs,'' posted at
{\tt http://info. phys.unm.edu/\verb|~|caves/reports/infopovm.pdf}.

\bibitem{King02}
C.~King, ``The Capacity of the Quantum Depolarizing Channel,'' {\tt
quant- ph/0204172}.

\bibitem{Caves02}
C.~M. Caves, C.~A. Fuchs and R.~Schack, ``Unknown Quantum States:\
The Quantum de Finetti Representation,'' J. Math.\ Phys. {\bf 43},
4537--4559 (2002); see also {\tt quant-ph/0104088}.

\bibitem{DAriano03}
G.~M. D'Ariano, P.~Perinotti, and M.~F. Sacchi, ``Informationally
Complete Measurements and Groups Representation,'' {\tt
quant-ph/0310013}.

\bibitem{WoottersPrivate}
W.~K. Wootters, private communication.

\bibitem{Barnum02}
H.~Barnum, ``Information-Disturbance Tradeoff in Quantum Measurement
on the Uniform Ensemble and on the Mutually Unbiased Bases,'' {\tt
quant-ph/0205155}.

\bibitem{Caves03}
See as an example, C.~M. Caves, I.~H. Deutsch, and R.~Blume-Kohout,
``Physical-Resource Demands for Scalable Quantum Computation,'' {\tt
quant-ph/0304083}.

\bibitem{Eotvos22}
R.~v. E\"otv\"os, D.~Pek\'ar, and E.~Fekete, ``Beitr\"age zum Gesetze
der Proportionalit\"at von Tr\"agheit und Gravit\"at,'' Annalen der
Physik {\bf 68}, 11--66 (1922).


\end{thebibliography}
\end{document}